\documentclass[jcs]{iosart2x}

\usepackage{amsmath,amssymb,amsfonts}
\usepackage{graphicx}
\usepackage{textcomp}
\usepackage{xcolor}  
\usepackage{algorithm,algcompatible}
\usepackage{listings}
\usepackage{makecell}
\usepackage{url}
\usepackage{multirow}

\definecolor{codegreen}{rgb}{0,0.6,0}
\definecolor{codegray}{rgb}{0.5,0.5,0.5}
\definecolor{codepurple}{rgb}{0.58,0,0.82}
\definecolor{backcolour}{rgb}{0.95,0.95,0.92}
\lstdefinestyle{mystyle}{
  backgroundcolor=\color{backcolour}, commentstyle=\color{codegreen},
  keywordstyle=\color{magenta},
  numberstyle=\tiny\color{codegray},
  stringstyle=\color{codepurple},
  basicstyle=\ttfamily\footnotesize,
  breakatwhitespace=false,         
  breaklines=true,   
  captionpos=b,      
  keepspaces=true,   
  numbers=left,      
  numbersep=4pt,    
  showspaces=false,  
  showstringspaces=false,
  showtabs=false,    
  tabsize=1,
  escapeinside={(*@}{@*)} 
}

\algnewcommand\algorithmicreturn{\textbf{return}}
\algnewcommand\RETURN{\algorithmicreturn}
\algnewcommand\algorithmicprocedure{\textbf{procedure}}
\algnewcommand\PROCEDURE{\item[\algorithmicprocedure]}%
\algnewcommand\algorithmicendprocedure{\textbf{end procedure}}
\algnewcommand\ENDPROCEDURE{\item[\algorithmicendprocedure]}%
\algnewcommand{\algvar}[1]{{\text{\ttfamily\detokenize{#1}}}}
\algnewcommand{\algarg}[1]{{\text{\ttfamily\itshape\detokenize{#1}}}}
\algnewcommand{\algproc}[1]{{\text{\ttfamily\detokenize{#1}}}}
\algnewcommand{\algassign}{\leftarrow}

\newcommand{\ie}{\textit{i.}\textit{e.}}

\def\BibTeX{{\rm B\kern-.05em{\sc i\kern-.025em b}\kern-.wanq08em
    T\kern-.1667em\lower.7ex\hbox{E}\kern-.125emX}}

\pubyear{0000}
\volume{0}
\firstpage{1}
\lastpage{1}

\begin{document}
\numberlinesfalse

\begin{frontmatter}


\title{HuntFUZZ: Enhancing Error Handling Testing through Clustering Based Fuzzing}


\runtitle{HuntFUZZ: Enhancing Error Handling Testing through Clustering Based Fuzzing}


\begin{aug}

\author[A,B]{\fnms{Jin} \snm{Wei}\ead[label=e1]{jwei17@fudan.edu.cn}}
\author[B,C]{\fnms{Ping} \snm{Chen}\ead[label=e2]{pchen@fudan.edu.cn}\thanks{Corresponding author. \printead{e2}.}}%
\author[D]{\fnms{Jun} \snm{Dai}\ead[label=e3]{}}
\author[D]{\fnms{Xiaoyan} \snm{Sun}\ead[label=e4]{}}
\author[A]{\fnms{Zhihao} \snm{Zhang}\ead[label=e5]{}}
\author[A]{\fnms{Chang} \snm{Xu}\ead[label=e6]{}}
\author[A]{\fnms{Yi} \snm{Wang}\ead[label=e7]{}}

\address[A]{School of Computer Science, \orgname{Fudan University}, Shanghai, \cny{China}\printead[presep={\\}]{e1}}
\address[B]{Institute of BigData, \orgname{Fudan University}, Shanghai, \cny{China}\printead[presep={\\}]{e2}}
\address[C]{Purple Mountain Laboratories, Nanjing, \cny{China}}
\address[D]{Worcester Polytechnic Institute, Massachusetts, \cny{USA}}

\end{aug}

\begin{abstract}
Testing a program's capability to effectively handling errors is a significant challenge, given that program errors are relatively uncommon.
To solve this, Software Fault Injection (SFI)-based fuzzing integrates SFI and traditional fuzzing, injecting and triggering errors for testing (error handling) code. 
However, we observe that current SFI-based fuzzing approaches have overlooked the correlation between paths housing error points. In fact, the execution paths of error points often share common paths. Nonetheless, Fuzzers usually generate test cases repeatedly to test error points on commonly traversed paths. This practice can compromise the efficiency of the fuzzer(s).
Thus, this paper introduces HuntFUZZ, a novel SFI-based fuzzing framework that addresses the issue of redundant testing of error points with correlated paths. 
Specifically, HuntFUZZ clusters these correlated error points and utilizes concolic execution to compute constraints only for common paths within each cluster. 
By doing so, we provide the fuzzer with efficient test cases to explore related error points with minimal redundancy.
We evaluate HuntFUZZ on a diverse set of 42 applications, and
HuntFUZZ successfully reveals 162 known bugs, with 62 of them being related to error handling. Additionally, due to its efficient error point detection method, HuntFUZZ discovers 7 unique zero-day bugs, which are all missed by existing fuzzers. Furthermore, we compare HuntFUZZ with 4 existing fuzzing approaches, including AFL, AFL++, AFLGo, and EH-FUZZ. 
Our evaluation confirms that HuntFUZZ can cover a broader range of error points, and it exhibits better performance in terms of bug finding speed.

\end{abstract}

\begin{keyword}
\kwd{Error Handling}
\kwd{Software Fault Injection}
\kwd{Fuzzing}
\kwd{Concolic Execution}
\kwd{Hybrid Fuzzing}
\end{keyword}

\end{frontmatter}


\section{Introduction} \label{Introduction}
Real-world programs necessitate error handling to address various potential errors that may occur. During program execution, exceptional situation may arise as a result of specific conditions, such as invalidate inputs, memory shortages, integer overflow, and network connection failures. These exceptional circumstances are typically referred to as errors, and the code responsible for handling these exceptions is known as error handling code.  
However, error handling within programs can often be problematic or entirely absent. Testing if a program can handle errors appropriately proves to be very challenging as this part of workflow is usually infrequently executed in normal programs, given the infrequency of errors~\cite{intro6,intro7,intro8}. The testing may also be inadequate at times due to the inherent difficulty of reaching error points during testing~\cite{intro1,intro2,intro3,intro4,intro5,intro6,intro7,intro8}. Insufficient error handling can lead to severe security consequences~\cite{severresult1, severresult2, severresult3, severresult4}, example vulnerabilities associated with error handling include CVE-2019-7846 causing information disclosure~\cite{CVE-2019-7846}, CVE-2019-2240 resulting in abnormal program functionality~\cite{CVE-2019-2240}, as well as CVE-2019-1750 and CVE-2019-1785 leading to DoS~\cite{CVE-2019-1750, CVE-2019-1785}. 
Thus, testing whether a program can handle errors is crucial to mitigate potential security risks.

To enhance the testing of error handling, recent solutions involve the utilization of Software Fault Injection (SFI)-based fuzzing approaches~\cite{POTUS,FIZZER,FIFUZZ,iFIZZ,EH-FUZZ}. These methods combine SFI~\cite{SFI} and fuzzing technologies~\cite{backgro1, backgro2, backgro3, backgro4, backgro5, backgro6, backgro7, backgro8, backgro9, backgro10, backgro11, backgro12, AFL, Honggfuzz, AFLFast, CollAFL, QSYM, REDQUEEN}.
Specifically, SFI introduces faults or errors into the tested program, so that the program can be executed to test whether it can effectively handle the injected faults or errors~\cite{FIFUZZ}. The code locations where errors are injected are referred to as error points. Fuzzing is then used to generate program inputs as test cases to cover paths leading to errors. Specifically, SFI-based fuzzing approaches require the mutation of error sequences to indicate where to insert errors at potential sites and mutation of program inputs to reach the error points.
Existing SFI-based fuzzing approaches primarily focus on how to generate error sequences. For instance, FIFUZZ~\cite{FIFUZZ} employs context-sensitive SFI to cover error points in different calling contexts. iFIZZ~\cite{iFIZZ} adopts a state-aware SFI approach, defining state as the runtime context of an error site, to reduce redundant fault scenarios. EH-FUZZ~\cite{EH-FUZZ} utilizes error coverage (a metric of a fuzzer's capability to test the number of injection scenarios for error points, which will be explained in details in Section~\ref{errorCoverage}) to direct fault injection, avoiding exploration of duplicate error scenarios and attempting to detect more diverse error circumstances. 

\begin{lstlisting}[language=C, emphstyle={\color{blue}}, caption={Patch for function \texttt{memalign} in \texttt{malloc.c} from GNU C Library 2.18 or earlier.},label={code:1}, escapeinside={(*@}{@*)}]
// Patch for function memalign
--- a/malloc/malloc.c
(*@\textcolor{red}{+++}@*) b/malloc/malloc.c
@@ -3015,6 +3015,13 @@ 
__libc_memalign(size_t alignment, size_t bytes)
{  ...
   if (alignment <  MINSIZE) alignment = MINSIZE;

(*@\textcolor{red}{+}@*)   /* Check for overflow. */ 
(*@\textcolor{red}{+}@*)  if (bytes > SIZE_MAX - alignment - MINSIZE)
(*@\textcolor{red}{+}@*)   {
(*@\textcolor{red}{+}@*)     __set_errno (ENOMEM);
(*@\textcolor{red}{+}@*)     return 0;
(*@\textcolor{red}{+}@*)    }
   arena_get(ar_ptr, bytes + alignment + MINSIZE);
   if(!ar_ptr)
      return 0;
}
\end{lstlisting}

Despite substantial efforts dedicated to generating error sequences, we find some insights that have not been considered in existing research on generating program inputs. Specifically, we observe that many error points exhibit significant correlation because they share common paths. However, existing fuzzers do not take this factor into account when generating program inputs, resulting in fuzzers repeatedly generating program inputs that cover these common paths.
For example, the CVE-2013-4332 vulnerability~\cite{CVE-2013-4332} indicates multiple integer overflows in \texttt{malloc.c} in GNU C Library version 2.18 and earlier. These vulnerabilities allow an attacker to cause integer overflow by manipulating the variable \texttt{bytes} in function \texttt{memalign}, \texttt{valloc} and \texttt{pvalloc}, leading to a denial of service (heap corruption). 
The vulnerability arises due to the lack of checks on the variable \texttt{bytes} in the program and the absence of capability of error handling when the value of \texttt{bytes} exceeds a certain range.
As shown in Code~\ref{code:1}, lines 9 -- 14 demonstrate the patch addressing the overflow vulnerability in the \texttt{memalign} function.
Specifically, a check is introduced in this patch to examine the variable \texttt{bytes}. If the value of this variable exceeds \texttt{SIZE\_MAX - alignment - MINSIZE}, an error message is thrown. Without this necessary check, integer overflow may happen. Similar overflow vulnerabilities also exist in the functions \texttt{valloc} and \texttt{pvalloc}, all due to the lack of checks on the variable \texttt{bytes}.
Furthermore, \texttt{memalign}, \texttt{valloc} and \texttt{pvalloc} reside within the same switch case structure and share a common call path, specifically ...$\rightarrow$ \texttt{allocate\_thread} $\rightarrow$ \texttt{allocate} $\rightarrow$ \texttt{allocate\_1} $\rightarrow$ \texttt{switch case} (as shown in Code~\ref{code:2}). This suggests that if a fuzzer separately solves the error points for function \texttt{memalign}, \texttt{valloc} and \texttt{pvalloc}, it may redundantly solve the common call path, thereby diminishing the efficiency of the fuzzer. We elaborate on this aspect in Section~\ref{insight}.

To enhance the capability of the SFI-based fuzzer in exploring error points, this paper proposes an optimization strategy that incorporates concolic execution to expedite the process of reaching related error points that share common paths. This strategy involves solving constraints only for the common path shared by these error points.
To implement this strategy, our approach starts by clustering error points that exhibit common paths. It ensures that within a given cluster, the distance between all error points and their common parent node is less than a specified threshold. 
Next, the significance of each cluster is assessed by calculating its weight. The weight of a cluster is determined by both the number of error points within the cluster and the distance of the cluster from the current execution path. This evaluation allows us to prioritize which cluster's path constraints should be computed by the concolic executor, focusing on the clusters that are most likely to lead to important and diverse error points.
Subsequently, we employ a constraint calculation approach to derive the constraints for the common path shared by the error points within each cluster. These constraints capture the conditions that must be satisfied for the execution to follow the common path leading to the error points. We then provide the fuzzer with test cases that satisfy these constraints, enabling it to efficiently explore and discover related error points with minimal redundant execution.

These strategies are implemented into a fuzzing framework named HuntFUZZ. We conduct a thorough experimental evaluation to validate its effectiveness and performance. 
The experimental results demonstrate that HuntFUZZ not only identifies known error handling bugs, but also discovers 7 zero-day bugs. Additionally, we compare HuntFUZZ with several state-of-the-art fuzzing methods, showing that it exhibits stronger error-point testing capabilities and broader error point coverage.

In conclusion, this paper makes the following contributions:

\begin{itemize}
\item We present HuntFUZZ, a novel SFI-based fuzzing framework that improves the efficiency of error point detection by employing clustering techniques. We also introduce an optimization algorithm that integrates concolic execution to effectively resolve input constraints for error points within each cluster. This approach not only enables targeted testing by directing the fuzzer towards specific error point clusters, but also prevents redundant exploration of error points sharing common paths. Consequently, it significantly enhances the fuzzer's effectiveness in error point testing. Our findings further demonstrate HuntFUZZ's capability in effectively exploring deep-state error points (as described in Section~\ref{cmp_bug}, which we define as error points with a depth exceeding 500 in the Control Flow Graph (CFG) in this paper). This is attributed to the integration of concolic execution, which aids in testing some deep-state error points dependent on very intricate and specific constraints.

\item We test HuntFUZZ across a diverse spectrum of 42 applications, including two datasets (Unibench~\cite{UniBench} and programs previously tested by EH-FUZZ~\cite{EH-FUZZ}) as well as 9 additional wild programs. HuntFUZZ discovered a total of 162 known bugs, including 62 error handling bugs. Additionally, HuntFUZZ uncovered 7 zero-day bugs. We compare it with four established fuzzing approaches (AFL~\cite{AFL}, AFL++~\cite{AFL++}, AFLGo~\cite{AFLGo}, and EH-FUZZ~\cite{EH-FUZZ}). The results affirm that HuntFUZZ can discover more error handling bugs and achieve accelerated and superior coverage of error points. Notably, compared to the contemporary SFI-based fuzzing method (\ie, EH-FUZZ), HuntFUZZ exhibits a remarkable 38.9\% increase in error coverage. 
\end{itemize}

\section{Background and Key Insights}
\subsection{Background}
\subsubsection{SFI-based Fuzzing for Error-handing Test}

Although errors in the program are not frequent, failure to handle errors properly can lead to serious security vulnerabilities, posing a significant threat to the normal operation of the system. Examples of such threats include denial of service, information disclosure, local privilege escalation, and other critical impacts~\cite{EH-FUZZ}.
While some traditional fuzzers~\cite{backgro1, backgro2, backgro3, backgro4, backgro5, backgro6, backgro7, backgro8, backgro9, backgro10, backgro11, backgro12, AFL, Honggfuzz, AFLFast, CollAFL, QSYM, REDQUEEN} are adept at discovering some errors by rapidly generating program inputs, these input-driven fuzzing approaches often fall short in detecting input-independent errors, because these types of errors typically stem from exceptional events that sporadically occur, such as insufficient memory or network connection failures. 
Thus, traditional fuzzers prove ineffective in handling errors due to its infrequent execution. 

To address the shortcomings of traditional fuzzers, researchers introduce SFI~\cite{SFI,sfifuzz1, sfifuzz3, sfifuzz4, sfifuzz5, sfifuzz6, sfifuzz7, sfifuzz8, intro5} into traditional fuzzing to trigger input-independent errors and force the execution of error paths. 
Specifically, SFI introduces faults or errors into the tested program and then runs the program to test whether it can effectively handle the injected faults or errors~\cite{FIFUZZ}. The code locations where errors are injected are referred to as error points. SFI-based fuzzing typically begins by conducting a static analysis of the source code of the tested program to identify error points. Subsequently, during each fuzzing loop, the fuzzer mutates error sequences, indicating whether the error points can execute normally or fail, and each error point includes the location and calling context of a covered error site~\cite{EH-FUZZ}. 
Then, SFI-based fuzzing approaches follow the traditional fuzzing procedure to generate and mutate program inputs based on code coverage. 
This fusion of SFI testing with fuzzing testing is known as SFI-based fuzzing~\cite{POTUS, FIZZER, FIFUZZ, iFIZZ, EH-FUZZ}. Among them, POTUS~\cite{POTUS} and FIZZER~\cite{FIZZER} focus on testing kernel-level drivers but overlook the execution contexts of injected faults and lack input mutation capabilities. iFIZZ~\cite{iFIZZ} targets IoT firmware applications, taking into account the execution contexts of injected faults, but lacks input mutation. FIFUZZ~\cite{FIFUZZ} considers the execution contexts of injected faults and supporting input mutation. It is designed for testing user-level applications. As the contemporary SFI-based fuzzing approach, EH-FUZZ~\cite{EH-FUZZ} can test both user-level applications and kernel-level modules, and it proposes using error coverage to guide the generation of error sequences.

\subsubsection{Concolic Execution and Hybrid Fuzzing} \label{back_2}
Concolic execution~\cite{SAGE,Driller,QSYM,KLEE,Mayhem,angr} is a software verification technique that combines concrete execution with traditional symbolic execution. In this approach, concrete inputs to the program are initially marked as symbolic variables. Then, concolic executor runs the target program according to a specific program input, collects constraints encountered during the execution path, and subsequently creates new program inputs by tracking these constraints.
The newly generated inputs are typically reintroduced into the system to investigate and explore various execution paths.

Traditional fuzzing is effective at rapidly generating program inputs, but it can only create inputs that lead to execution paths with loose branch conditions~\cite{Hybridfuzztesting}. In contrast, concolic execution excels at discovering inputs that lead to execution paths with complex branch conditions~\cite{QSYM}. 
To take advantage of traditional fuzzing and concolic execution, a hybrid approach, known as hybrid fuzzing~\cite{HybridConcolicTesting,Hybridfuzztesting,Driller,QSYM,SymCC}, was proposed. In hybrid fuzzing, the concolic executor receives program inputs from the fuzzer and generates new program inputs, potentially enabling the exploration of new execution paths. This assists the fuzzer in uncovering paths with complex branch conditions.

\subsection{Key Insights} \label{insight}

By analyzing locations of error points, we observe a notable correlation among the paths to error points. In particular, many error points share common paths from the program's entry point to the occurrence of the error. This suggests that when the fuzzer separately addresses these error points, it may redundantly traverse the common call paths, potentially diminishing the efficiency of the fuzzer. However, this issue is currently overlooked in existing SFI-based fuzzing methods.
For example, in Section~\ref{Introduction}, we discussed the vulnerability CVE-2013-4332 in the GNU C Library, which leads to integer overflows in the functions \texttt{memalign}, \texttt{valloc}, and \texttt{pvalloc}. The vulnerability in each function arises from manipulating the variable \texttt{bytes} in a way that causes it to exceed the maximum representable value for the integer data type. Hence, we consider each manipulation of the variable \texttt{bytes} in these functions \texttt{memalign}, \texttt{valloc}, and \texttt{pvalloc} as an exploitable error point.
Next, we consider the path relationship of the functions \texttt{memalign}, \texttt{valloc}, and \texttt{pvalloc}. As shown in Code~\ref{code:2}, these three functions are within the same \texttt{switch case} structure. This \texttt{switch case} structure is invoked by the function \texttt{allocate\_1}, and based on the value of the variable \texttt{allocation\_function}, it selects one of the functions \texttt{memalign}, \texttt{valloc}, or \texttt{pvalloc} to execute. Therefore, within this \texttt{switch case} structure, there are three error points that need to be tested, occurring at Line 8, Line 13, and Line 18. Additionally, the calling path for these three error points is common, traversing through … $\rightarrow$ \texttt{allocatethread} $\rightarrow$ \texttt{allocate} $\rightarrow$ \texttt{allocate\_1} $\rightarrow$ \texttt{switch case}. If we use a fuzzer to individually explore these three error points, it would require generating test cases repeatedly to execute along each path. 
In this paper, we aim to minimize the redundancy in exploring these paths. We strategically cluster these error points and leverage concolic execution to compute the input constraints for the common path leading to these three error points. Subsequently, the fuzzer only needs to vary values in the program input minimally to reach different error points. For instance, in Code~\ref{code:2}, changing the value of the variable \texttt{allocation\_function} would be sufficient.

\begin{lstlisting}[language=C, emphstyle={\color{blue}}, caption={\texttt{memalign}, \texttt{valloc} and \texttt{pvalloc} reside within the same switch case structure.},label={code:2}]
// ... -> allocate_thread -> allocate -> allocate_1 -> switch case
allocate_1 (void)
{
    switch (allocation_function)
    {
        case with_memalign:
      {
        void *p = memalign (alignment, allocation_size); // error point 1
        return (struct allocate_result) {p, alignment};
      }
        case with_valloc:
      {
        void *p = valloc (allocation_size); // error point 2
        return (struct allocate_result) {p, page_size};
      }
        case with_pvalloc:
      {
        void *p = pvalloc (allocation_size); // error point 3
        return (struct allocate_result) {p, page_size};
      }    
    }
}
\end{lstlisting}

Through this clustering strategy, several benefits can be enabled:
\begin{itemize}
    \item \textbf{Improved effectiveness in testing error points for SFI-based fuzzing methods.} By reducing the redundant exploration of common paths among error points, HuntFUZZ can test more error sequences within the same timeframe compared to existing SFI-based fuzzing methods. We validate this conclusion in Section~\ref{errorCoverage}. Furthermore, to verify the effect of clustering, we compare the number of error sequences tested with versus without clustering in Section~\ref{cluster_k}, finding that the clustering method indeed helps test more error sequences effectively.
    \item \textbf{Enhanced detection of deep-state error points.} Existing SFI-based fuzzing methods rely on traditional fuzzing methods to generate program inputs, which may struggle to test some deep-state error points dependent on very intricate and specific constraints~\cite{QSYM}, as discussed in Section~\ref{back_2}. However, HuntFUZZ utilizes concolic execution to strategically compute input constraints within a cluster, which may include deep-state error points. This systematic approach assists the fuzzer in covering such deep-state error points more comprehensively. This is validated in Section~\ref{cmp_bug}.
\end{itemize}

\section{Design of HuntFUZZ} 

\begin{figure}[htbp]
 \centering
\includegraphics[width=0.7\linewidth]{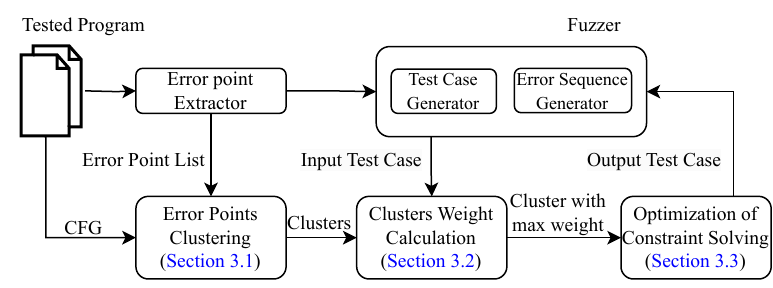}
\caption{Framework of HuntFUZZ.}
\label{fig:1}
\end{figure}

In this section, we explain our design of HuntFUZZ.
The overall architecture is illustrated in Fig.~\ref{fig:1}. 
Firstly, HuntFUZZ staticly analyzes the tested program to extract error points using the error point extractor.
Like the existing SFI-based fuzzers~\cite{EH-FUZZ}, the fuzzer's test case generator then produces program inputs for executing the target program, following a traditional fuzzing approach.
The fuzzer also has an error sequence generator to determine the execution status of error points (covered or not) based on their calling context. Consequently, it generates error sequences indicating whether the error points can execute normally (indicated as 0) or fail (indicated as 1) due to an injected fault.
Meanwhile, the fuzzer also gathers runtime information and detects bugs.

In addition to the general flow described above, this paper innovatively introduces the following three extra modules:
\begin{itemize}
    \item Error Points Clustering: This process involves clustering all error points based on the Control Flow Graph (CFG) of the tested program and the error point list (generated by error point extractor). Error points grouped into the same cluster typically share the common path. Moreover, error points within the same cluster tend to have distances from their nearest common parent node that fall within a specific range. We will elaborate on this aspect in Section~\ref{Error Points Clustering}.
    \item Cluster Weight Calculation: In this process, HuntFUZZ receives test cases generated by the fuzzer and calculates the weight of each cluster based on the number of error points in each cluster and the distance from the current execution path which is determined by the input test case. The cluster with the highest weight is then selected as the cluster that the fuzzer aims to reach. We will elaborate on this process in Section~\ref{Cluster Weight Calculation}.
    \item Optimization of Constraint Solving: To avoid redundant computation of constraints for each cluster of error points, we propose an optimization approach, specifically, computing constraints only for the longest common path of each cluster. The concolic executor then provides the fuzzer with a test case capable of reaching the common parent node for each cluster of error points. This process continues until the fuzzer generates test cases reaching all error points in this cluster or until the number of generated test cases reaches a predefined threshold (indicating potential difficulty in reaching the error points). In such cases, the concolic executor recalculates input constraints for the next cluster that needs coverage. We will elaborate on this aspect in Section~\ref{Optimization Constraint Solving}.
\end{itemize}


\textbf{Technical Challenges.} We pinpoint three challenges in implementing our approach: 1) How to design a clustering method for error points? 2) How to calculate the weight for each cluster?
3) How to design an optimization algorithm to efficiently compute the input constraints for the longest common path of error points in a cluster?

\begin{figure}[htbp]
 \centering
\includegraphics[width=0.45\linewidth]{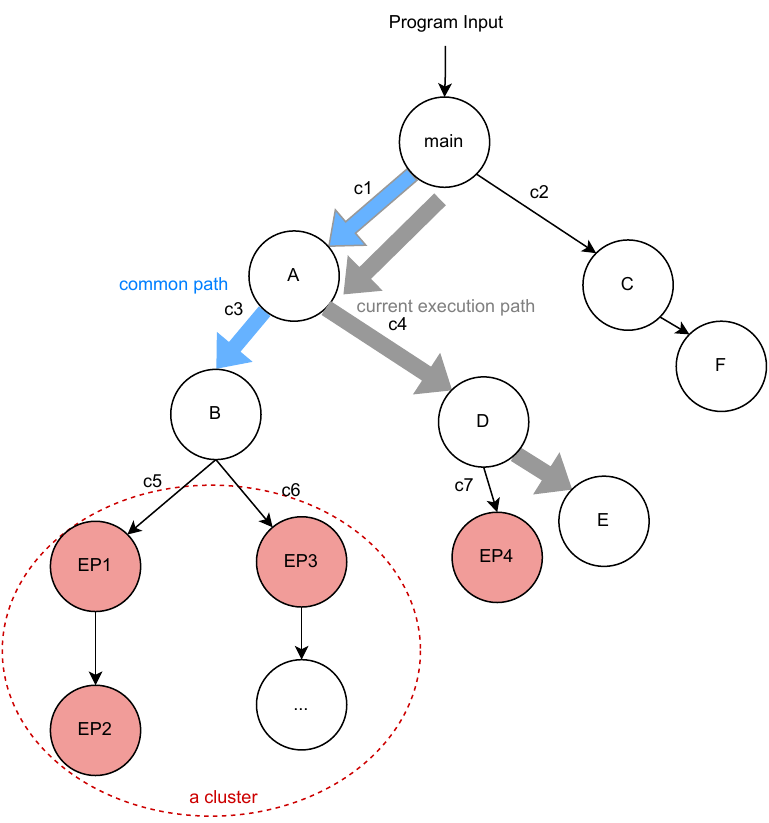}
\caption{A CFG of a tested program along with error points that need testing.}
\label{fig:CFG}
\end{figure}

To illustrate our idea, we provide an example as shown in Fig.~\ref{fig:CFG}. We schematically illustrate a partial CFG of a tested program along with error points that need testing. Nodes in the CFG represent basic blocks of the program, so nodes EP1 to EP4 represent the basic blocks where the four tested error points reside. Edges in the CFG represent constraints on program inputs to reach that node. The program's entry point is the \texttt{main} function, which takes 
program input and executes specific program paths. Suppose a certain program input leads to the program following the path: \texttt{main} $\rightarrow$ \texttt{A} $\rightarrow$ \texttt{D} $\rightarrow$ \texttt{E}, and we want the fuzzer to test error points along other paths. 
For instance, intuitively, error points EP1, EP2, and EP3, which are closely located and share a common path: \texttt{main} $\rightarrow$ \texttt{A} $\rightarrow$ \texttt{B}. To effectively reach these three error points, we desire the concolic executor to output a test case reaching the common parent node B, allowing the fuzzer to then generate separate test cases for each error point. Given the proximity of node B to these three error points, we consider the fuzzer can easily reach them. 
In addition to guiding the fuzzer to reach these three error points, this approach also offers the advantage of efficiency, as it reduces redundancy in exploration for the common path repetitively for each error point.



\subsection{Error Points Clustering} \label{Error Points Clustering}
To identify error points with common paths, in this module, we propose a clustering algorithm for error points. Moreover, for a given $k$ value, this algorithm ensures that the error points grouped together have a distance from their nearest common parent node that is less than or equal to $k$.

\begin{algorithm} 
\footnotesize
\caption{Error Points Clustering} 
\label{alg:1}
\begin{algorithmic}[1]
    \REQUIRE error point list $E$, control flow graph $G$
    \ENSURE error point clauster $EPCs$
    \PROCEDURE{getErrorPointCluster} ($E$,$G$) 
    \STATE $EPCs \gets [~]$
    \STATE $S \gets  [~]$
    \STATE $bbkSet \gets  [~]$
    \STATE // Step 1
    \FOR {$i \gets 0$ to $\text{length}(E)$}
    \STATE $bbk \gets \text{getFatherList}(E[i], G, k)$
    \STATE $bbkSet[i] \gets bbk$
    \STATE $S[i] \gets \text{false}$
    \ENDFOR
    \STATE // Step 2
    \WHILE{$\text{existUnvisited}(S)$}
    \STATE $CEI \gets \text{getRandomEP}(E, S)$
    \STATE $S[CEI] \gets \text{true}$
    \STATE $P \gets [~]$    
    \FOR{$i \gets 0$ to $\text{length}(bbkSet)$}
      \IF {$\text{isSamePATH}(S[CEI], S[i])$}
        \STATE $P.\text{add}(E[i])$
        \STATE $S[i] \gets \text{true}$
      \ELSE 
         \IF {$\text{hasCommon}(bbkSet[CEI], bbkSet[i])$}
        \STATE $P.\text{add}(E[i])$
        \STATE $S[i] \gets \text{true}$
      \ENDIF
      \ENDIF
    \ENDFOR    
    \STATE $EPCs.\text{add}(P)$
  \ENDWHILE  
  \STATE \textbf{return} $EPCs$
  \ENDPROCEDURE
\end{algorithmic} 
\end{algorithm}

As illustrated in Algorithm~\ref{alg:1}, in Step 1, for each error point, we traverse upward for $k$ iterations to get a set of $k$ nodes. This set is referred to as \texttt{bbkSet}. In Step 2, we compare the \texttt{bbkSets} of error points. If there are $n$ ($n$ $\geq$ 2) error points sharing a common node within their respective \texttt{bbkSets}, it signifies that the common ancestor's distance from these $n$ error points is less than or equal to $k$. Consequently, these $n$ error points are clustered together. If this condition is not met, clustering cannot be performed. 
For instance, in Fig.~\ref{fig:CFG}, for the error points EP1, EP2, EP3 and EP4, when $k=2$, the \texttt{bbkSets} for these four error points are respectively: \{B, A\}, \{EP1, B\}, \{B, A\}, \{D, A\}. It can be observed that EP1, EP2, and EP3 share a common parent node B, and the distance of these three error points to B is less than or equal to 2. Therefore, we can cluster EP1, EP2, and EP3 into one group. It's worth noting that EP1, EP3, and EP4 also have a common parent node A. However, our algorithm chooses to prioritize clustering EP1 and EP2, which belong to the same path (as shown in Line 16 -- Line 18 of Algorithm~\ref{alg:1}), because such error points often share longer common paths. For instance, the common path for \{EP1, EP2, EP3\} is \texttt{main} $\rightarrow$ \texttt{A} $\rightarrow$ \texttt{B}, while \{EP1, EP3, EP4\} shares the common path \texttt{main} $\rightarrow$ \texttt{A}. Clearly, the first clustering method results in error points with longer common paths. Besides, reducing the redundant exploration of longer common paths implies a greater improvement in the efficiency of the fuzzer.
As a result, the final clustering result for these four error points is set1: \{EP1, EP2, EP3\}, set2: \{EP4\}.

\subsection{Cluster Weight Calculation} \label{Cluster Weight Calculation}
The concolic executor executes specific paths based on test cases generated by fuzzer. However, we need the concolic executor to guide the fuzzer in covering error points within a specific cluster. To determine which cluster to cover, we propose a strategy that prioritizes covering clusters with a higher number of error points and clusters that are closer to the current execution path.

To achieve this goal, as shown in Algorithm~\ref{alg:2}, we calculate the weight of each cluster by factoring in the number of points within the cluster and their distance from the current execution path. More precisely, we conduct a weighted sum of these two metrics, ultimately identifying the cluster with the highest weight (Line 5 in Algorithm~\ref{alg:2}).

\begin{algorithm}
\footnotesize
\caption{Get Cluster Weight} 
\label{alg:2} 
\begin{algorithmic}[1]
    \REQUIRE error point clusters $EPCs$, test case generated by fuzzer $FI$
    \ENSURE $clusterWeight$
    \PROCEDURE{getClusterWeight} ($EPCs$, $FI$) 
    \STATE $maxCw \gets 0$
    \FOR{$i \gets 0$ \textbf{to} \text{length}$(EPC)$}
    \STATE $EPNum \gets \text{getEPNum}(EPC[i])$ 
    \STATE $clusterDistance \gets \text{getClusterDistance}(EPC[i], FI)$ 
    \STATE $clusterWeight \gets w1 \times EPNum + w2 \times clusterDistance$
    \IF{$clusterWeight > maxCw$}
        \STATE $maxCw \gets clusterWeight$
    \ENDIF
    \ENDFOR
    \STATE \textbf{return} $maxCw$
  \ENDPROCEDURE
\end{algorithmic} 
\end{algorithm}

\subsection{Optimization of Constraint Solving} \label{Optimization Constraint Solving}
In this module, HuntFUZZ is designed to optimize the selection and resolution of specific constraint conditions, specifically, HuntFUZZ strategically focuses on clustered error points along the common path. By doing so, the concolic executor only returns the constraints relevant to the common paths of these error points. For example, regarding the clustered error points EP1, EP2, and EP3 in Fig.~\ref{fig:CFG}, the concolic executor calculates the common path constraints for these three error points, \ie, \{c1, c3\}, and provides fuzzer test cases satisfying these constraints.

We define the criteria for determining the completion of cluster detection as whether the fuzzer is covering the currently injected error points within the cluster or if the number of test cases generated by the fuzzer exceeds a certain threshold. In the latter case, we may consider that some error points within the cluster are too challenging to be covered.
During this cluster detection process, the concolic executor receives test cases generated by the fuzzer. Based on the weight of error point cluster, the concolic executor calculates the constraints of common paths of this cluster's error points. The concolic executor then outputs a test case that satisfies these constraints to fuzzer. Subsequently, the fuzzer continuously generates inputs until the completion of this cluster detection.

Therefore, the optimization algorithm of the concolic executor is depicted in Algorithm~\ref{alg:3}.
For Line 1 -- Line 9, we initialize certain parameters. We initialize a CFG ($G$) based on the binary program. Additionally, we initialize the error point list ($E$), error point path ($TP$), and cluster error points ($EPCs$) according to Algorithm~\ref{alg:1}. Simultaneously, we initialize the count of inputs generated by the current fuzzer ($curMutationCount$), the cluster of maximum weight ($maxCw$, based on Algorithm~\ref{alg:2}) and the $constraints$ generated by the concolic executor. 
Subsequently, Line 10 -- Line 29 constitute the main flow of the concolic executor. In this process, the concolic executor continuously receives program inputs generated by the fuzzer and sends the test cases reaching the error points back to the fuzzer. Within this flow, Line 11 -- Line 18 indicate that if all error points in a cluster are covered or if the fuzzer-generated inputs reach $mutateThreshold$ (indicating that generating inputs to cover the error point is deemed challenging), then constraints for the next cluster are regenerated.
Line 19 -- Line 29 signify the verification of whether the fuzzer-generated input covers the error point. If coverage is achieved, the error point coverage status is updated. If not, the concolic executor proceeds to receive further inputs ($FI$) to locate coverage for the error point.

\begin{algorithm}
\footnotesize
\caption{Optimization Algorithm of Concolic Executor} 
\label{alg:3} 
\begin{algorithmic}[1]
    \REQUIRE tested program  $P$, all error-points location info $EP$, test case generated by fuzzer $FI$
    \ENSURE test case satisfying $constraint$
    \STATE $G \gets \text{getCFG}(P)$
    \STATE $E \gets \text{getErrorPointList}(EP, ES)$ 
    \STATE $TP \gets \text{getErrorPointPath}(ES, G, EP)$ 
    \STATE $EPCs \gets \text{getErrorPointCluster}(E, G)$ 
    \STATE $curMutationCount \gets 0$ 
    \STATE $maxCw \gets \text{getClusterWeight}(EPCs,FI)$
    \STATE $constraint \gets \text{getConstraint}(maxCw)$
    \STATE \textbf{calculate} $constraint$ 
    \STATE \textbf{send} $test case$ based on $constraint$ \textbf{to Fuzzer}
    \WHILE{$(FI)$}
        \IF{$curMutationCount > \text{mutateThreshold}$ \textbf{or} $clusterState == covered$}
            \STATE $nextCw \gets \text{getClusterWeight}(EPCs,FI)$
            \STATE $constraint \gets \text{getConstraint}(nextCw)$
            \STATE $curMutationCount \gets 0$
            \STATE \textbf{calculate} $constraint$ 
            \STATE \textbf{send}  $test case$ based on $constraint$ \textbf{to Fuzzer}
            \STATE \textbf{continue}
        \ENDIF
        \STATE $curMutationCount \gets curMutationCount + 1$
        \STATE $coverErrorFlag \gets \text{false}$
        \FOR{$i \gets 0$ \textbf{to} $\text{length}(TP)$}
            \IF{$FI == TP[i]$}
                \STATE $coverErrorFlag \gets \text{true}$
            \ENDIF
        \ENDFOR
        \IF{$coverErrorFlag == \text{true}$}
            \STATE \textbf{continue}
        \ENDIF
    \ENDWHILE
\end{algorithmic} 
\end{algorithm}

\section{Implementation}
In this section, we elaborate on the details of implementing HuntFUZZ, covering four main aspects: error points extractor, code instrumentation, SFI-based fuzzer, and concolic executor.

\textbf{\emph{Error Points Extractor.}} Our approach extracts function calls as error points, as recent studies~\cite{EH-FUZZ,ep2,FIFUZZ} indicate that the majority of error points involve code statements checking error-indicating return values of function calls. We identify candidate error points by examining functions that return pointers or integers.
Additionally, we employ 9 distinct exception-handling methods to aid in the recognition of error handling functions. These methods include 4 categories implemented through unconditional branching statements, including \texttt{return}, \texttt{break}, \texttt{continue}, and \texttt{goto}. The other 5 categories involve custom functions handling exceptional states, such as logging (\texttt{log}), program termination (\texttt{exit}), closing files or directories (\texttt{close}), deleting files or directories (\texttt{delete}), and freeing memory (\texttt{free}).

\textbf{\emph{Code Instrumentation.}}
We instrument three types of code snippets at compilation time:
1) to record the runtime context of each error point execution, we insert monitoring code at the entry and exit points of each called function;
2) for monitoring the execution states of error points and conducting fault injection at runtime, we instrument code before each identified error point. If the value of an error point in the error sequence is 1, a fault is injected into that specific error point;
3) additionally, we instrument basic blocks in code to collect code coverage during the execution process of the fuzzer.

\textbf{\emph{SFI-based fuzzer.}}
The fuzzer follows established practices in the literature, such as the approach of EH-FUZZ~\cite{EH-FUZZ}. The error sequence generator creates error sequences to determine whether to inject faults into the identified error points. The input generator mutates and generates new program inputs. Besides, we use bug checkers, such as ASan~\cite{Asan} and MSan~\cite{Msan}, to analyze runtime information about memory to determine if it may trigger bugs.

\textbf{\emph{Concolic Executor.}}
We draw inspiration from QSYM~\cite{QSYM} for the implementation of the concolic executor. The concolic executor is deployed on the Intel Pin~\cite{PIN} tool. Additionally, we modify the constraint solving component of QSYM, incorporating 1.2K lines of code for our clusters weight calculation and optimization constraint solving algorithm. Furthermore, 0.7K lines of code are added to facilitate communication with fuzeer for implementing relevant system calls.

\section{Evaluation}
In this section, we evaluate the following questions:
\begin{itemize}
    \item How effective is HuntFUZZ in discovering bugs in real-world applications? Can it discover zero-day bugs?
    \item How does HuntFUZZ perform compared to other state-of-the-art fuzzing approaches in terms of found bugs, error coverage and code coverage?
    \item How do the parameters associated with the algorithm(s) influence the overarching efficacy of the HuntFUZZ framework?    
\end{itemize}

\textbf{Experimental Environment and Setup.}
We conduct our experiments on a machine powered by an Intel(R) Xeon(R) Gold 5118 CPU @ 2.30GHz with 16 cores. The experiments are performed on an Ubuntu 20.04.5 LTS operating system.  
To validate HuntFUZZ, we evaluate it on two datasets (UniBench~\cite{UniBench} and applications previously tested by EH-FUZZ~\cite{EH-FUZZ}), as well as 9 wild applications. UniBench comprises 20 test applications, while EH-FUZZ consists of 15 test applications. Due to the duplication of two applications across these two datasets, we conduct a total of 42 application tests.
The basic information of these applications is listed in the Table~\ref{tab:1}.

\textbf{Error Point Identification.}
For the tested applications, we first utilize HuntFUZZ to statically analyze their source code, identifying potential error points. Subsequently, we manually identify realistic error points capable of causing failures and errors. Table~\ref{tab:1} displays the error points recognized by HuntFUZZ. Overall, HuntFUZZ identifies 18213 error points. Among them, we manually confirm these error points and ultimately determine 10684 realistic error points. Indeed, the manual selection of realistic error sites is not challenging, as many error sites call the same functions. Checking each error site does not require a significant amount of effort.

\begin{table}[t]
\centering
\scriptsize
\caption{\textbf{Information about the applications tested with HuntFUZZ, as well as the identified error points and bug information found by HuntFUZZ.}}
\resizebox{\linewidth}{!}{
\begin{tabular}{cccccccc}
\hline
& Tested program     & Version           & Identified error points & Realistic error points & Known bugs & Error handling bugs & Zero-day bugs \\ 
\hline
\multirow{20}{*}{Unibench~\cite{UniBench}} & exiv2              & 0.26              & 136                                                                & 70                                                               & 3          & 2                   &   0            \\
& gdk-pixbuf-pixdata & gdk-pixbuf 2.31.1 & 107                                                                & 63                                                               & 2          & 0                   &    0           \\
& Jasper             & 2.0.12+2.0.14            & 227                                                                & 92                                                               & 4          & 2                   &    1           \\
& jhead              & 3.00              & 530                                                                & 359                                                              & 5          & 3                   &    0           \\
& libtiff            & 3.9.7+4.5.1             & 985                                                                & 695                                                              & 7          & 3                   &      1         \\
& lame               & 3.99.5            & 830                                                                & 332                                                              & 5          & 1                   &    0           \\
& mp3gain            & 1.5.2             & 273                                                                & 198                                                              & 1          & 0                   &    0           \\
& swftools           & 0.9.2             & 674                                                                & 571                                                              & 3          & 0                   &     0          \\
& ffmpeg             & 4.0.1             & 198                                                                & 112                                                              & 13         & 7                   &     0          \\
& flvmeta            & 1.2.1             & 636                                                                & 254                                                              & 2          & 0                   &    0           \\
& Bento4             & 1.5.1-628         & 581                                                                & 348                                                              & 6          & 2                   &     0          \\
& cflow              & 1.6 + 1.7               & 117                                                                & 88                                                               & 1          & 0                   &     0          \\
& ncurses            & 6.1               & 525                                                                & 210                                                              & 3          & 0                   &    0           \\
& jq                 & 1.5               & 618                                                                & 485                                                              & 6          & 2                   &    0           \\
& mujs               & 1.0.2             & 431                                                                & 279                                                              & 2          & 0                   &     0          \\
& pdftotext          & 4.00              & 328                                                                & 165                                                              & 3          & 1                   &    0           \\
& SQLite             & 3.8.9             & 153                                                                & 91                                                               & 5          & 3                   &    0           \\
& binutils           & 2.28              & 362                                                                & 144                                                              & 6          & 2                   &     0          \\
& libpcap            & 1.8.1             & 731                                                                & 329                                                              & 4          & 1                   &    0           \\
& tcpdump            & 4.8.1             & 912                                                                & 626                                                              & 9          & 3                   &    0           \\ \hline
\multirow{13}{*}{EH-FUZZ~\cite{EH-FUZZ}} & vim                & 8.2.3595         & 334                                                                & 270                                                              & 5          & 2                   &    0           \\
& bison              & 3.8.1            & 187                                                                & 125                                                              & 0          & 0                   &    0           \\
& nasm               & 2.15.05          & 62                                                                 & 26                                                               & 0          & 0                   &    0           \\
& catdoc             & 0.95             & 101                                                                & 69                                                               & 5          & 4                   &    0           \\
& clamav             & 0.104.1          & 2125                                                               & 1247                                                             & 3          & 1                   &    0           \\
& gif2png+libpng     & 2.5.14+1.6.3    & 129                                                                & 65                                                               & 0          & 0                   &    0           \\
& OpenSSL            & 3.0.0+3.0.9            & 135                                                                & 102                                                              & 4          & 3                   &    1          \\
& btrfs              & Linux 5.16.16     & 929                                                                & 351                                                              & 3          & 1                   &    0           \\
& xfs                & Linux 5.16.16     & 201                                                                & 171                                                              & 1          & 1                   &    0           \\
& jfs                & Linux 5.16.16     & 114                                                                & 100                                                              & 2          & 1                   &    0           \\
& cephfs             & Linux 5.16.16     & 460                                                                & 140                                                              & 4          & 3                   &    0           \\
& xhci               & Linux 5.16.16     & 180                                                                & 104                                                              & 1          & 1                   &     0          \\
& vmxnet3            & Linux 5.16.16     & 98                                                                 & 43                                                               & 3          & 1                   &   0\\ \hline

\multirow{9}{*}{wild program} & man-db~\cite{man-db} & 2.12.0 & 295 & 158 & 5 & 2 & 0\\
& woff2~\cite{woff2} & 1.0.2 & 163 & 139 & 3 & 0 & 0\\
& gzip~\cite{gzip} & 1.13 & 397 & 272 & 6 & 2 & 0\\
& bzip2~\cite{bzip2} & 1.0.6  & 432 & 365 & 5 & 2 & 0\\
& sassc~\cite{sassc} & 3.6.2 & 321 & 284  & 3 & 0 & 0\\
& tidy~\cite{tidy} & 5.9.20 & 527 & 381 & 2 & 0 & 1 \\
& jqlang~\cite{jqlang} & 1.7 & 118 & 67 & 3 & 1 &  1\\
& bash~\cite{bash} & 5.2.21 & 827 & 351 & 8 & 3 &  1\\
& mksh~\cite{mksh} & mksh-R59c & 724  & 343 & 6 & 2 &  1\\

\hline
\bf{total}   &  42 & & 18213 & 10684 & 162 & 62 & 7
\\\hline
\end{tabular}}

\label{tab:1}
\end{table}

\subsection{Found Bugs}
For the error points indicated in Table~\ref{tab:1}, we utilize HuntFUZZ to conduct testing upon them. HuntFUZZ tests each program using ASan~\cite{Asan} and MSan~\cite{Msan} to detect bugs, limiting the testing time to 24 hours, and repeat the experiment 5 times. The results of bugs we found are shown in Table~\ref{tab:1}. 
Overall, HuntFUZZ has discovered 162 known bugs, among which 62 are related to error handling.
For bugs leading to program crashes or failures, we manually examine their root causes using bug reports and source code to determine whether they are known bugs or unique zero-day bugs. Notably, HuntFUZZ discovered 7 zero-day bugs in \texttt{Jasper}, \texttt{libtiff}, \texttt{OpenSSL}, \texttt{tidy}, \texttt{jqlang}, \texttt{bash} and \texttt{mksh}. We have responsibly reported these zero-day bugs. The zero-day bug in \texttt{libtiff} has been confirmed by the developers. The zero-day bug in \texttt{jqlang} is also simultaneously found by OSS-FUZZ~\cite{OSS-FUZZ}, and this bug has already been fixed.
We are awaiting responses regarding the other bugs.

Here, we provide a detailed overview of the zero-day bugs discovered in \texttt{Jasper}, \texttt{libtiff}, and \texttt{OpenSSL}. The detailed information about the other zero-day bugs is shown in Appendix~\ref{other_0_day}.

\textbf{\emph{Wild free bug in Jasper.}} In Code~\ref{code:Jasper}, the function \texttt{jas\_iccprof\_create} makes use of \texttt{jas\_malloc} (Line 5) and \texttt{jas\_iccattrtab\_create} (Line 9) to check whether the variables \texttt{prof} and \texttt{prof->attrtab} are allocated correctly, with both \texttt{jas\_malloc} and \texttt{jas\_iccattrtab\\\_create} encapsulating the \texttt{malloc} function. At this point, upon detecting that \texttt{prof} is successfully allocated (\texttt{prof} $\neq$ NULL) while \texttt{prof->attrtab} allocation fails (\texttt{prof->attrtab} = NULL), the program proceeds to Line 14, entering error handling. Consequently, Lines 11 and 12 are not executed, leaving the variable \texttt{prof->tagtab.ents} uninitialized. However, the error handling code (Lines 14 -- Line 17) invokes \texttt{jas\_iccprof\_destroy}, and due to the uninitialized \texttt{prof->tagtab.ents}, when attempting to free \texttt{prof->tagtab.ents} (Line 25), a wild free bug occurs.

\textbf{\emph{NULL-pointer dereference bug in libtiff.}}
In Code~\ref{code:3}, within the function \texttt{TIFFReadDirectory}, there is an \texttt{if} statement that checks whether the return value of the function \texttt{\_TIFFMergeFieldInfo} is NULL (Line 4). The second parameter of \texttt{\_TIFFMergeFieldInfo} is the return value of \texttt{\_TIFFCreateAnonFieldInfo}. The function \texttt{\_TIFFCreateAnonFieldInfo} uses \texttt{\_TIFFmal\\-loc} to allocate memory for the variables \texttt{fld} and \texttt{fld->field\_name} (Line 24 and Line 27). When there is a failure in allocating memory for either of these variables, the return value of \texttt{\_TIFFCreateAnonFieldInfo} is NULL. In such a scenario, calling \texttt{\_TIFFCreateAnonFieldI\\-nfo} (Line 4) leads to a NULL pointer dereference bug.

\vspace{7pt}

\begin{lstlisting}[language=C, emphstyle={\color{blue}}, caption={Wild free bug in \texttt{Jasper}.},label={code:Jasper}, escapeinside={(*@}{@*)}]
// jas_iccprof_create -> jas_iccprof_destroy -> jas_free

static jas_iccprof_t *jas_iccprof_create()
{   
    if (!(prof = jas_malloc(sizeof(jas_iccprof_t)))) 
    {
        goto error;
    }
    if (!(prof->attrtab = jas_iccattrtab_create()))
        goto error;
    prof->tagtab.numents = 0;
    prof->tagtab.ents = 0;
    return prof;
error:
    if (prof)
        jas_iccprof_destroy(prof);
    return 0;
}

void jas_iccprof_destroy(jas_iccprof_t *prof)
{
    if (prof->attrtab)
        jas_iccattrtab_destroy(prof->attrtab);
    if (prof->tagtab.ents)
        jas_free(prof->tagtab.ents); 
    jas_free(prof);
}

void jas_free(void *ptr)
{
    free(ptr);
}
\end{lstlisting}

\vspace{-15pt}
\begin{lstlisting}[language=C, emphstyle={\color{blue}}, caption={NULL-pointer dereference bug in \texttt{libtiff}.},label={code:3}, escapeinside={(*@}{@*)}]
// TIFFReadDirectory -> _TIFFMergeFieldInfo -> _TIFFCreateAnonFieldInfo -> ...-> _TIFFmalloc
TIFFReadDirectory(TIFF* tif)
{ 
    if (!_TIFFMergeFieldInfo(tif, _TIFFCreateAnonFieldInfo(tif, dp->tdir_tag, (TIFFDataType) dp->tdir_type),1)) 
}

int _TIFFMergeFieldInfo(TIFF* tif, const TIFFFieldInfo info[], int n)
{
    for (i = 0; i < n; i++)
    {
        const TIFFFieldInfo *fip =
            _TIFFFindFieldInfo(tif, info[i].field_tag, info[i].field_type);

        if (!fip) {
            *tp++ = (TIFFFieldInfo*) (info + i);
            tif->tif_nfields++;
        }
    }    
    return n;
}

TIFFFieldInfo* _TIFFCreateAnonFieldInfo(...)
{   
    fld = (TIFFFieldInfo *) _TIFFmalloc(sizeof (TIFFFieldInfo));
    if (fld == NULL)
        return NULL;
    fld->field_name = (char *) _TIFFmalloc(32);
    if (fld->field_name == NULL) 
    {
        return NULL;
    }
}

void* _TIFFmalloc(tsize_t s)
{
    return (malloc((size_t) s));
}
\end{lstlisting}

\vspace{-10pt}
\textbf{\emph{NULL-pointer dereference bug in OpenSSL.}}
When testing the OpenSSL custom module X509 with insufficiently allocated space for X509, it can result in the function \texttt{do\_cmd} calling the function \texttt{lh\_FUNCTION\_retrieve}, which sets the value of the variable \texttt{fp} to NULL (Line 5, Code 5). Subsequently, when invoking the function \texttt{EVP\_get\_digestbyname} in Line 8, it leads to the execution of the function \texttt{ossl\_lib\_ctx\_get\_data} (Line 1). Function \texttt{ossl\_lib\_ctx\_get\_data} is responsible for retrieving context information (Line 15) and dereferencing the variable \texttt{ctx->lock} (Line 16). Besides, the function \texttt{context\_init} initializes the structure variable \texttt{ctx}. When initialization fails, the error handling code is executed, setting all fields of \texttt{ctx} to NULL. This results in a NULL-pointer dereference bug when dereferencing the variable \texttt{ctx->lock} in Line 16.

\vspace{-8pt}
\begin{lstlisting}[language=C, emphstyle={\color{blue}}, caption={NULL-pointer dereference bug in \texttt{OpenSSL}.},label={code:4}]
// do_cmd -> lh_FUNCTION_retrieve -> EVP_get_digestbyname ->...-> ossl_lib_ctx_get_data
static int do_cmd()
{  
    //fp: retrieve function pointer
    fp = lh_FUNCTION_retrieve(prog, &f); 
    if (fp == NULL) 
    {
        if (EVP_get_digestbyname(argv[0])) {...}
    }
    return 1;
}

void *ossl_lib_ctx_get_data()
{   
    ctx = ossl_lib_ctx_get_concrete(ctx);     
    if (!CRYPTO_THREAD_read_lock(ctx->lock)) 
        return NULL;
}

static int context_init(OSSL_LIB_CTX *ctx)
{   
    ctx->oncelock = CRYPTO_THREAD_lock_new();
    if (ctx->oncelock == NULL)
        goto err;
    return 1;
 err:
    memset(ctx, '\0', sizeof(*ctx));
    return 0;
}
\end{lstlisting}

\textbf{Bug Features.} 
We attribute HuntFUZZ's ability to discover zero-day bugs to its capacity to achieve higher error coverage than other fuzzers within the same timeframe (refer to Section~\ref{errorCoverage}). 
Error coverage signifies the fuzzer's proficiency in thoroughly testing scenarios involving the injection of errors.
Reviewing these zero-day bugs, three key observations emerge: 
1) firstly, most of errors caused by these bugs revolve around operations on pointer-type data. For example, \texttt{Jasper}'s bug involves a wild free operation on an uninitialized pointer, while the other two involve dereference operations on NULL pointers. This suggests that incorrect operations on pointers are prone to triggering program crashes or failures;
2) secondly, we find that \texttt{Jasper} and \texttt{OpenSSL} bugs result from incorrect error handling functionality, while the bug in \texttt{libtiff} is caused by a failed \texttt{malloc} operation. This indicates that our tool can detect not only bugs related to error handling but also other types of bugs leading to program crashes or failures;
3) thirdly, we find that some zero-day bugs require the simultaneous activation of two error points to trigger. For example, for \texttt{Jasper}, there are two error points: one where \texttt{prof} = \texttt{jas\_malloc(sizeof(jas\_iccprof\_t))} (Line 5 in Code~\ref{code:Jasper}), and the other where \texttt{prof->attrtab} $\neq$ \texttt{jas\_iccattrtab\_create()} (Line 9 in Code~\ref{code:Jasper}). The bug in \texttt{libtiff} has two error points: \texttt{fld} = NULL (Line 24 in Code~\ref{code:3}) and \texttt{fld->field\_name} = NULL (Line 27 in Code~\ref{code:3}). The OpenSSL bug also has two error points: \texttt{fp} = NULL (Line 5 in Code~\ref{code:4}) and \texttt{ctx->lock} = NULL (Line 25 in Code~\ref{code:4}). 

\subsection{Comparison to Existing Fuzzing Approaches}
We select four state-of-the-art fuzzing approaches for comparison on testing 33 applications from two datasets (Unibench~\cite{UniBench} and applications tested by EH-FUZZ~\cite{EH-FUZZ}), including three traditional fuzzers (AFL~\cite{AFL}, AFL++~\cite{AFL++} and AFLGo~\cite{AFLGo}) and one SFI-based fuzzer: EH-FUZZ~\cite{EH-FUZZ}. 
It is worth to note that within the current landscape of SFI-based fuzzing approaches~\cite{POTUS, FIZZER, FIFUZZ, iFIZZ, EH-FUZZ}, both POTUS~\cite{POTUS} and iFIZZ~\cite{iFIZZ} are limited to testing specific domains of applications. Specifically, POTUS is tailored for USB drivers testing, while iFIZZ is designed for testing of IoT firmware applications. Since these tools do not align with the applications we intend to test, and to our knowledge, FIZZER~\cite{FIZZER} and FIFUZZ~\cite{FIFUZZ} are not yet open source, we ultimately opt for EH-FUZZ as the tool for comparison with HuntFUZZ.
We compare HuntFUZZ with selected/representative fuzzing tools in terms of found bugs, error coverage (the number of covered error sequences) and code coverage (the number of covered code branches). For HuntFUZZ, we configure the parameter values associated with the optimization algorithm at their default value, \ie, $k = 2$, $w1 = w2 = 0.5$, and $mutateThreshold = 10,000$.

\begin{table}[t]
\centering
\scriptsize
\caption{\textbf{The results of comparing HuntFUZZ with four state-of-the-art fuzzing approaches in terms of found bugs, error coverage, and code coverage.}}
\resizebox{\linewidth}{!}{

\begin{tabular}{c|cc|cc|cc|ccccc|ccccc}
\hline
                                 & \multicolumn{2}{c|}{AFL}                               & \multicolumn{2}{c|}{AFL++}                             & \multicolumn{2}{c|}{AFLGo}                             & \multicolumn{5}{c|}{EH-FUZZ}                                                                                                                & \multicolumn{5}{c}{HuntFUZZ}                                                                                                               \\ \cline{2-17} 
                                 &                       &                               &                       &                               &                       &                               &                                           & \multicolumn{2}{c}{Depth}           &                               &                          &                                           & \multicolumn{2}{c}{Depth}           &                               &                          \\ \cline{9-10} \cline{14-15}
\multirow{-3}{*}{Tested Program} & \multirow{-2}{*}{Bug} & \multirow{-2}{*}{Branch} & \multirow{-2}{*}{Bug} & \multirow{-2}{*}{Branch} & \multirow{-2}{*}{Bug} & \multirow{-2}{*}{Branch} & \multirow{-2}{*}{\begin{tabular}[c] {@{}c@{}} Bug \\ (error handling bug) \end{tabular}} & \textless{}500 & \textgreater{}=500 & \multirow{-2}{*}{Branch} & \multirow{-2}{*}{ErrSeq} & \multirow{-2}{*}{\begin{tabular}[c] {@{}c@{}} Bug \\ (error handling bug) \end{tabular}} & \textless{}500 & \textgreater{}=500 & \multirow{-2}{*}{Branch} & \multirow{-2}{*}{ErrSeq} \\ \hline
exiv2                            & 0                     & 5636                          & 0                     & 17497                         & 0                     & 12945                         & 1(1)                                      & 1              & 0                  & 26484                         & 23741                    & 3(2)                                      & 1              & 1                  & 11353                         & 35372                    \\
gdk\_pixbuf\_pixdata               & 0                     & 9256                          & 0                     & 14536                         & 0                     & 14823                         & 0                                         & 0              & 0                  & 17256                         & 44841                    & 2(0)                                      & 0              & 0                  & 7821                          & 58952                    \\
jasper                           & 0                     & 7904                          & 0                     & 11253                         & 0                     & 12553                         & 2(1)                                      & 1              & 0                  & 15904                         & 5684                     & 4(2)                                      & 1              & 1                  & 6978                          & 7023                     \\
jhead                            & 0                     & 9953                          & 0                     & 9952                          & 0                     & 10234                         & 2(1)                                      & 1              & 0                  & 15564                         & 7831                     & 5(3)                                      & 1              & 2                  & 8342                          & 9102                     \\
libtiff                          & 0                     & 4246                          & 0                     & 29547                         & 2                     & 11568                         & 3(1)                                      & 1              & 0                  & 36850                         & 30495                    & 7(3)                                      & 2              & 1                  & 14109                         & 44051                    \\
lame                             & 0                     & 5242                          & 0                     & 26234                         & 0                     & 19835                         & 1(0)                                      & 0              & 0                  & 27230                         & 961                      & 5(1)                                      & 0              & 1                  & 15967                         & 1237                     \\
mp3gain                          & 0                     & 7254                          & 0                     & 10632                         & 0                     & 9894                          & 0                                         & 0              & 0                  & 17546                         & 213                      & 1(0)                                      & 0              & 0                  & 9448                          & 1149                     \\
swftools                         & 0                     & 6351                          & 0                     & 16569                         & 0                     & 13056                         & 0                                         & 0              & 0                  & 21654                         & 24510                    & 3(0)                                      & 0              & 0                  & 12305                         & 36294                    \\
ffmpeg                           & 0                     & 5127                          & 2                     & 13246                         & 3                     & 12065                         & 10(5)                                     & 4              & 1                  & 25312                         & 12368                    & 13(7)                                     & 4              & 3                  & 10320                         & 19376                    \\
flvmeta                          & 0                     & 4465                          & 0                     & 14254                         & 0                     & 10934                         & 1(0)                                      & 0              & 0                  & 16446                         & 2984                     & 2(0)                                      & 0              & 0                  & 8295                          & 4208                     \\
Bento4                           & 0                     & 9962                          & 0                     & 22156                         & 0                     & 22105                         & 2(1)                                      & 1              & 0                  & 24304                         & 5692                     & 6(2)                                      & 1              & 1                  & 17730                         & 8843                     \\
cflow                            & 0                     & 2258                          & 0                     & 15061                         & 0                     & 15964                         & 1(1)                                      & 1              & 0                  & 16218                         & 367                      & 1(0)                                      & 0              & 0                  & 4692                          & 992                      \\
ncurses                          & 0                     & 3433                          & 0                     & 23972                         & 0                     & 19127                         & 1(1)                                      & 1              & 0                  & 28338                         & 2072                     & 3(0)                                      & 0              & 0                  & 19934                         & 3015                     \\
jq                               & 0                     & 4549                          & 0                     & 34501                         & 0                     & 12257                         & 4(2)                                      & 2              & 0                  & 35554                         & 3591                     & 6(2)                                      & 2              & 0                  & 12413                         & 4218                     \\
mujs                             & 0                     & 2542                          & 0                     & 8839                          & 0                     & 10213                         & 1(0)                                      & 0              & 0                  & 12240                         & 1536                     & 2(0)                                      & 0              & 0                  & 6491                          & 2059                     \\
pdftotext                        & 0                     & 3807                          & 0                     & 19956                         & 0                     & 14247                         & 1(1)                                      & 1              & 0                  & 28723                         & 2985                     & 3(1)                                      & 1              & 0                  & 14302                         & 3645                     \\
SQLite                           & 0                     & 5154                          & 0                     & 27562                         & 0                     & 23395                         & 4(2)                                      & 2              & 0                  & 31154                         & 3051                     & 5(3)                                      & 2              & 1                  & 19934                         & 4959                     \\
binutils                         & 0                     & 4450                          & 0                     & 16749                         & 0                     & 19452                         & 3(2)                                      & 2              & 0                  & 24601                         & 3774                     & 6(2)                                      & 2              & 0                  & 10556                         & 4095                     \\
libpcap                          & 0                     & 3924                          & 0                     & 12985                         & 0                     & 19576                         & 3(1)                                      & 1              & 0                  & 13024                         & 3352                     & 4(1)                                      & 1              & 0                  & 9560                          & 3857                     \\
tcpdump                          & 0                     & 3123                          & 0                     & 20121                         & 2                     & 18755                         & 7(3)                                      & 3              & 0                  & 24702                         & 1976                     & 9(3)                                      & 3              & 0                  & 11345                         & 2975                     \\
vim                              & 0                     & 5937                          & 0                     & 19305                         & 0                     & 12958                         & 3(1)                                      & 1              & 0                  & 28317                         & 23968                    & 5(2)                                      & 1              & 1                  & 10851                         & 35524                    \\
bison                            & 0                     & 4085                          & 0                     & 14521                         & 0                     & 11975                         & 0                                         & 0              & 0                  & 17022                         & 11956                    & 0                                         & 0              & 0                  & 10835                         & 43877                    \\
nasm                             & 0                     & 4366                          & 0                     & 7834                          & 0                     & 8347                          & 0                                         & 0              & 0                  & 10118                         & 1285                     & 0                                         & 0              & 0                  & 6431                          & 3102                     \\
catdoc                           & 1                     & 586                           & 2                     & 675                           & 0                     & 663                           & 2(1)                                      & 1              & 0                  & 1998                          & 821                      & 5(4)                                      & 1              & 3                  & 759                           & 1070                     \\
clamav                           & 0                     & 6961                          & 0                     & 17140                         & 0                     & 13125                         & 3(1)                                      & 1              & 0                  & 19903                         & 13124                    & 3(1)                                      & 1              & 0                  & 12145                         & 16832                    \\
gif2png+libpng                   & 0                     & 5167                          & 0                     & 4246                          & 0                     & 3452                          & 0                                         & 0              & 0                  & 7123                          & 53                       & 0                                         & 0              & 0                  & 3245                          & 192                      \\
openssl                          & 0                     & 7835                          & 0                     & 12249                         & 1                     & 13125                         & 3(1)                                      & 1              & 0                  & 26484                         & 24536                    & 4(3)                                      & 1              & 2                  & 8240                          & 37684                    \\
btrfs     & -                     &      -                         & -                     &     -                          & -                     &   -                            & 2(1)                                      & 1              & 0                  & 11235                         & 894                      & 3(1)                                      & 1              & 0                  & 1207                          & 1052                     \\
xfs      & -                     &   -                            & -                     &   -                            & -                     &    -                           & 0                                         & 0              & 0                  & 23845                         & 1042                     & 1(1)                                      & 0              & 1                  & 3481                          & 2154                     \\
jfs      & -                     &    -                           & -                     &    -                           & -                     &  -                             & 1(1)                                      & 1              & 0                  & 8459                          & 1230                     & 2(3)                                      & 1              & 2                  & 2895                          & 2665                     \\
cephfs   & -                     &      -                         & -                     &   -                            & -                     &    -                           & 3(2)                                      & 2              & 0                  & 12395                         & 739                      & 4(1)                                      & 0              & 1                  & 1968                          & 1549                     \\
 xhci      & -                     &      -                         & -                     &    -                           & -                     &      -                         & 0                                         & 0              & 0                  & 4292                          & 1293                     & 1(1)                                      & 0              & 1                  & 3863                          & 2705                     \\
vmxnet3   & -                     & -                             & -                     &    -                           & -                     &   -                            & 2(1)                                      & 1              & 0                  & 2153                          & 1346                     & 3(1)                                      & 1              & 0                  & 4017                          & 3723                     \\ \hline 
total                            & 1                     & 143573                        & 4                     & 441592                        & 8                     & 366643                        & 66(32)                                    & 31(96.9\%)     & 1(3.1\%)           & 632448                        & 264311                   & 121(50)                                   & 28(56\%)       & 22(44\%)           & 301832                        & 407551   \\    \hline           
\end{tabular}}

\label{tab:compare}
\end{table}

\subsubsection{Comparison on Found Bugs} \label{cmp_bug}
Due to the fact that AFL, AFL++, and AFLGo are only capable of testing user-level applications, they are utilized to assess the user-level applications listed in Table~\ref{tab:compare}. It is observed that AFL++ and AFLGo outperforms AFL in discovering more bugs, owing to their integration of superior strategies for input mutation and seed selection. 
However, due to the absence of injection error points in these three fuzzing methods, they face challenges in detecting bugs related to error handling. Throughout our testing process, these three tools do not identify bugs associated with error handling.

Compared to the aforementioned three fuzzers, EH-FUZZ~\cite{EH-FUZZ} and HuntFUZZ both have the capability to test kernel-level applications. Overall, for the user-level and kernel-level applications listed in Table~\ref{tab:compare}, HuntFUZZ has demonstrated the discovery of a greater number of bugs compared to EH-FUZZ, particularly in the realm of error handling bugs. Moreover, HuntFUZZ identifies all the error handling bugs detected by EH-FUZZ.

In addition, regarding EH-FUZZ and HuntFUZZ, 
We summarize the depths of error points which trigger error handling bugs in the CFG. We find that out of the 32 error handling bugs discovered by EH-FUZZ, depths of 31 bugs' error points < 500. Conversely, among the 50 error handling bugs found by HuntFUZZ, depths of 22 bugs' error points $\geq$ 500. This finding demonstrates that HuntFUZZ has the ability to test error points with deeper depth. We believe this is because for some deep-state error points, the program inputs must adhere to very intricate and specific constraints. EH-FUZZ, using traditional fuzzing methods to generate inputs, may struggle to test these deep-state error points (as we discussed in Section~\ref{back_2}). In contrast, HuntFUZZ leverages concolic execution to purposefully compute input constraints within a cluster, which can include deep-state error points. This helps the fuzzer systematically cover such deep-state error points.

\subsubsection{Comparison on Error Coverage} \label{errorCoverage}

\begin{figure}[h]
 \centering
\includegraphics[width=\textwidth]{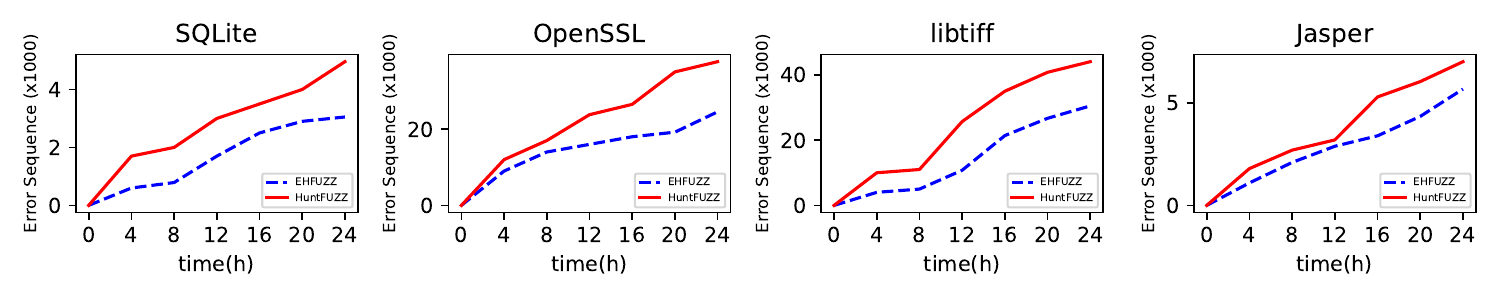}
\caption{Comparsion of HuntFUZZ and EH-FUZZ in terms of error coverage.}
\label{fig:cmp_ec}
\end{figure}

Since the three traditional fuzzing approaches (AFL, AFL++ and AFLGo) cannot conduct fault injection, we compare HuntFUZZ with the representative SFI-based fuzzing method (EH-FUZZ) in terms of error coverage. The results are shown in Table~\ref{tab:compare}.
Similar to EH-FUZZ~\cite{EH-FUZZ}, in this paper, error coverage represents the number of error sequences (indicating whether the error points can execute normally or fail). The ability to cover more error sequences signifies that the fuzzer can test more scenarios where errors are injected.
As shown in Fig.~\ref{fig:cmp_ec}, we select four applications—\texttt{SQLite}, \texttt{OpenSSL}, \texttt{libtiff}, and \texttt{Jasper}—to showcase the number of error sequences tested by both EH-FUZZ and HuntFUZZ. 
We can see that HuntFUZZ exhibits higher error coverage compared to EH-FUZZ. This is because HuntFUZZ utilizes
clustering strategy, assisting the fuzzer achieving superior error coverage at a faster pace. Fig.~\ref{fig:cmp_ec} shows that HuntFUZZ rapidly ramps up to reach error points, surpassing EH-FUZZ. 
For example, when testing \texttt{SQLite}, HuntFUZZ can test approximately 2,000 error sequences within about 8 hours. In contrast, EH-FUZZ takes around 16 hours to test the same number of error sequences (HuntFUZZ has an efficiency improvement of roughly double).
Notably, averaging across experiments spanning 24 hours for each program test, HuntFuzz achieves 38.9\% higher than EH-FUZZ.

\subsubsection{Comparison on Code Coverage}

\begin{figure}[h]
 \centering
\includegraphics[width=\textwidth]{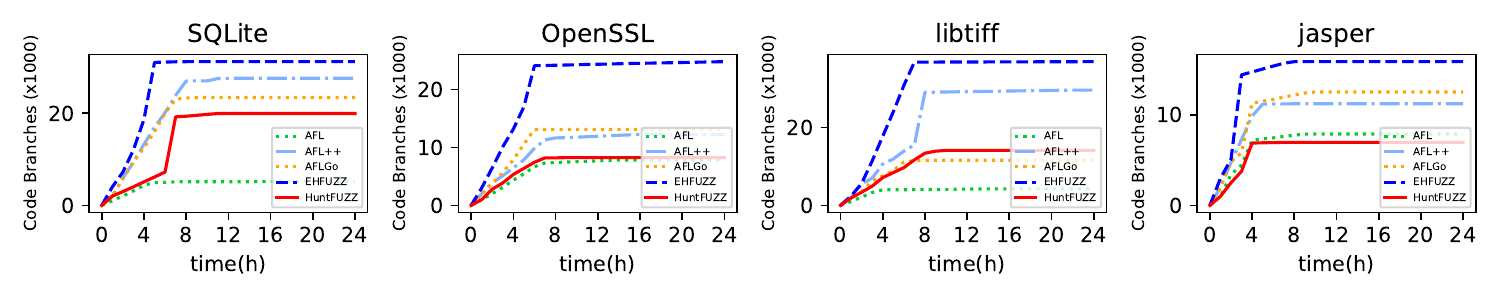}
\caption{Comparsion of HuntFUZZ, AFL, AFL++, AFLGo, and EH-FUZZ in terms of code coverage.}
\label{fig:cmp_cc}
\end{figure}
For code coverage, we compare this metric by summarizing the number of code branches tested by various fuzzers. As shown in Table~\ref{tab:compare}, for the tested applications, HuntFUZZ does not achieve the highest code coverage. This is because HuntFUZZ is more focused on the code branches where error points reside, rather than concerning itself with branches that do not contain error points. As illustrated in the Fig.~\ref{fig:cmp_cc}, we select four applications—\texttt{SQLite}, \texttt{OpenSSL}, \texttt{libtiff}, and \texttt{Jasper}—to demonstrate how code branches evolve over time under the influence of five fuzzing approaches. In general, EH-FUZZ tends to exhibit higher code coverage compared to other three traditional fuzzing approaches. This is attributed to EH-FUZZ covering error points in different calling contexts, a feature that drives it to explore more code branches. However, EH-FUZZ may explore some code branches that are unrelated to error points.
On the other hand, HuntFUZZ does not show a significant increase in code coverage compared to other tools. This is because HuntFUZZ does not need to consider branches unrelated to errors but rather aims for higher error coverage.

\subsection{The Impact of Parameters in Algorithms}\label{subsec:parameters}
In this section, we explore the impact of several parameters of optimization algorithm on the error coverage of HuntFUZZ . We conduct these experiments on 7 applications, including \texttt{SQLite}, \texttt{OpenSSL}, \texttt{libtiff}, \texttt{Jasper}, \texttt{jhead}, \texttt{ffmpeg} and \texttt{libpcap}. These parameters include the distance parameter $k$ in the error points clustering algorithm (Algorithm~\ref{alg:1}), the weighted metrics $w1$ and $w2$ in the cluster weight calculation algorithm (Algorithm~\ref{alg:2}), and the threshold $mutateThreshold$ for the number of test cases generated by the fuzzer in the optimization algorithm (Algorithm~\ref{alg:3}).
When investigating the impact of a specific parameter on the error coverage of HuntFUZZ, we maintain the values of the other variables at their defaults. We define the default values for these three parameters as follows: $k=2$, $w1=w2=0.5$, and $mutateThreshold=10,000$.

\begin{figure}[h]
 \centering
\includegraphics[width=0.9\textwidth]{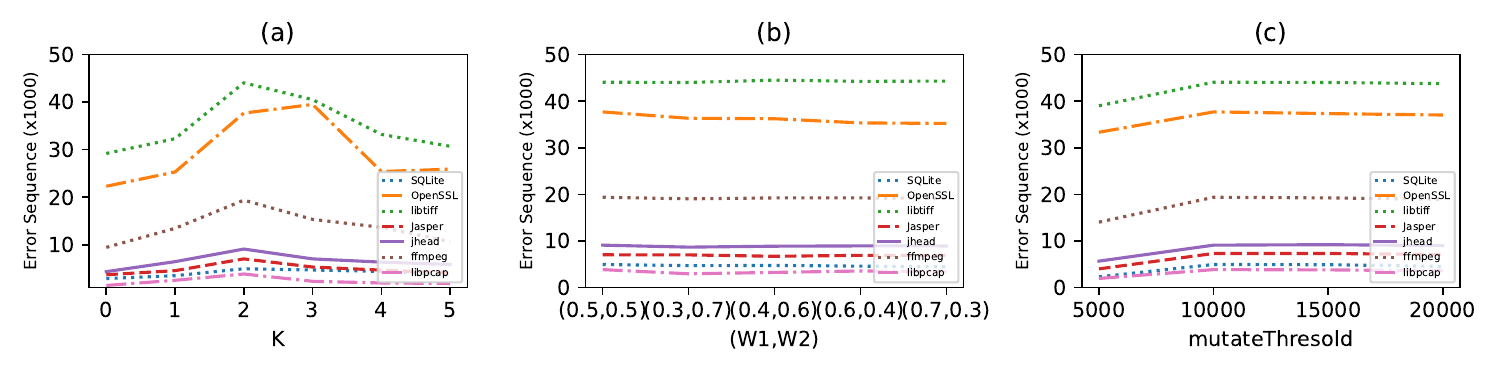}
\caption{The influence of $k$, $w1,w2$, and $mutateThreshold$ to error coverage.}
\label{fig:impact}
\end{figure}

\subsubsection{Cluster Distance $k$} \label{cluster_k}

In Algorithm~\ref{alg:1}, the parameter $k$ signifies that the distance of error points within a cluster to their common parent node is less than or equal to $k$. Consequently, for a given tested program, the value of $k$ influences the number of error points within a cluster. As depicted in Fig.~\ref{fig:impact}(a), we illustrate the impact of varying $k$ values on the error coverage of HuntFUZZ across 7 applications. And our results represent the error coverage observed after conducting tests on these applications for 24 hours.
When $k = 0$, it implies that the distance from the error point to the common parent is 0,
meaning that there is no actual clustering. In this scenario, the concolic executor needs to compute input constraints separately for each error point, leading to redundant calculations of constraints for common paths of error points. As the value of $k$ increases, the number of error points within a cluster grows, allowing the concolic executor to calculate constraints only for the common paths of these error points. Consequently, the performance of the concolic executor improves. However, the concolic executor can only guide the fuzzer to the common parent node of these error points, requiring the fuzzer to still attempt coverage of error points within the cluster. Therefore, as $k$ continues to increase, the performance of the fuzzer decreases. Hence, during the same testing duration, both excessively small and overly large values of $k$ can adversely impact the error coverage of HuntFUZZ.

\subsubsection{Cluster Weights $w1$ and $w2$}

In Algorithm~\ref{alg:2}, $w1$ and $w2$ are weights assigned to the parameters $EPNum$ (number of error points) and $clusterDistance$ (distance between the cluster and the current execution path) when calculating the cluster weight. These weights signify the importance of $EPNum$ and $clusterDistance$ in determining the weight of a cluster.
In Fig.~\ref{fig:impact}(b), we present the impact of different values for $w1$ and $w2$ on the error coverage of HuntFUZZ. 
It can be observed that different values of $w1$ and $w2$ lead to slight variations in error coverage. In general, for the majority of applications, when $w1=0.5$ and $w2=0.5$, the error coverage of HuntFUZZ is maximized after 24 hours.

\subsubsection{mutateThreshold}

In Algorithm~\ref{alg:3}, when the fuzzer attempts to cover an error point, the $mutateThreshold$ signifies the point at which the exploration of this error point stops once the fuzzer generates a specified number of test cases. In Fig.~\ref{fig:impact}(c), we document the impact of different $mutateThreshold$ values on the error coverage of HuntFUZZ. It is evident that as the $mutateThreshold$ value increases, for instance, from 5000 to 10,000, there is an improvement in HuntFUZZ's error coverage after 24 hours. However, when the $mutateThreshold$ value becomes excessively large (such as 20000), the error coverage almost plateaus and may even exhibit a slight decline. This is attributed to the fact that an excessively large $mutateThreshold$ consumes too much time on that specific error point, hindering the exploration of other error points and causing the overall error coverage to stabilize or slightly decrease.


\section{Related Work}
Many recent studies~\cite{POTUS,FIZZER,FIFUZZ,iFIZZ,EH-FUZZ} have utilized SFI-based fuzzing to trigger infrequently-executed errors in programs, covering various scenarios such as USB drivers~\cite{POTUS}, device drivers~\cite{FIZZER}, and IoT firmware~\cite{iFIZZ}. These techniques typically mutate both error sequences and program inputs together, aiming to test whether error points will trigger error handling bugs. However, a common challenge in SFI-based fuzzing is the issue of early crash, where the execution stops if an error is encountered, preventing the testing from reaching deep error paths.
To address this challenge, FIFUZZ~\cite{FIFUZZ} introduces a context-sensitive error injection method that effectively distinguishes shallow and deep error points, thus avoiding injecting shallow errors when testing deep error points. iFIZZ~\cite{iFIZZ} tackles the problem by saving the context (error's call stack) of error points to prevent the reproduction of previously tested error points. 
On the other hand, EH-FUZZ~\cite{EH-FUZZ} argues that using code coverage to guide error sequence generation is unreasonable since if two test cases trigger the same error point but in different execution contexts, these methods would consider them as equivalent. However, the contexts in which these error points are triggered may differ, and code coverage cannot reflect the context information of error points. In light of this, EH-FUZZ proposes using error coverage to guide the generation of error sequences. Error sequences consist of the execution status (failure or execute normally) of error points and their context information. This approach allows for a more comprehensive testing of handling errors by considering the diverse contexts in which errors can occur, rather than relying solely on code coverage-based guidance.

However, existing SFI-based fuzzing relies on traditional fuzzing for test case generation, 
these approaches do not consider the correlation of paths where error points are located. This leads to fuzzers needing to repeatedly generate test cases to reach duplicated paths, thereby diminishing the efficiency of the fuzzer.
This paper introduces HuntFUZZ, which addresses the aforementioned limitations in SFI-based fuzzing by incorporating concolic execution. 
Taking into account the correlation among paths where certain error points are situated, HuntFUZZ only computes constraints for common paths, thereby enhancing the efficiency of the fuzzer in exploring these error points.

\section{Conclusion}
In this paper, we introduce HuntFUZZ, by considering correlations among paths containing error points and selectively computing constraints for common paths. Specifically, we propose an algorithm for clustering error points with common paths, calculating the weight of each cluster, and utilizing an optimization strategy to explore clusters with the highest weights. HuntFUZZ surpasses current SFI-based fuzzing methods with faster and superior error coverage, specifically showing a substantial 38.9\% increase compared to the most advanced SFI-based fuzzing method. Moreover, HuntFUZZ detects zero-day bugs that other tools failed to find.

Furthermore, although we observe the correlation of error points' paths, such path correlations may be prevalent across the fuzzer's targets beyond error handling scenarios. In addition to error handling scenarios, more general contexts may also benefit from clustering targets to reduce fuzzers' exploration of redundant paths. We will delve deeper into this in future work.







\nocite{*}
\bibliographystyle{ios1}           
\bibliography{jcs_template
}        

%

\appendix

\section{Other Zero-day Bugs Found by HuntFUZZ}
\label{other_0_day}
In this section, we present zero-day bugs discovered by HuntFUZZ in other applications.

\textbf{\emph{READ memory access bug in tidy.}}
As shown in Code~\ref{code:tidy}, within the \texttt{InsertDocType} function, there is a while loop with the condition \texttt{!nodeIsHTML(element)} (Line 6). The expanded condition of the \texttt{nodeIsHTML} macro checks if \texttt{element} exists, if \texttt{element->tag} exists, and if \texttt{element->tag->id} is equal to a specific value \texttt{tid} (Line 1 -- Line 2).
The bug arises when \texttt{element} is NULL. In this case, the condition \texttt{!nodeIsHTML(element)} will evaluate to \texttt{true}, causing the loop body to be executed.
Inside the loop body, an attempt is made to access a NULL pointer, specifically \texttt{element->parent} (Line 7). This results in a ``READ memory access'' error, as attempting to read from a NULL pointer is invalid.

\begin{lstlisting}[language=C, emphstyle={\color{blue}}, caption={READ memory access bug in \texttt{tidy}.},label={code:tidy}]
#define TagIsId(node, tid) ((node) && (node)->tag && (node)->tag->id == tid)
#define nodeIsHTML( node ) TagIsId( node, tidyTag_HTML )

static void InsertDocType( tidyDocImpl* doc, Node *element, Node *doctype )
{   ...
    while ( !nodeIsHTML(element) )
        element = element->parent;
    ...
}
\end{lstlisting}

\textbf{\emph{Heap overflow bug in jqlang.}}
As shown in Code~\ref{code:jqlang}, in the function \texttt{jvp\_literal\_number\_liter\\-al}, a length \texttt{len} is calculated as \texttt{jvp\_dec\_number\_ptr(n)->digits + 14} (Line 4). Then, a memory buffer of length \texttt{len} is allocated, and the pointer to this buffer is stored in \texttt{plit->literal\_data} using the \texttt{jv\_mem\_alloc} function (Line 5). Subsequently, the \texttt{decNumbe\\-rToString(pdec, plit->literal\_data)} function is called to convert \texttt{pdec} into a string and store it in the memory buffer pointed by \texttt{plit->literal\_data} (Line 6). However, the \texttt{decNumberToString} function internally calculates the length of the string as \texttt{len + 15}. The memory allocation in the \texttt{jvp\_literal\_number\_literal} function only allocates a buffer of length \texttt{len} without considering the additional length.
Therefore, when \texttt{decNumberToString} writes the string to the memory buffer pointed by \texttt{plit->literal\_data}, it may exceed the allocated memory buffer's range, resulting in a heap overflow bug.

\begin{lstlisting}[language=C, emphstyle={\color{blue}}, caption={Heap overflow bug in \texttt{jqlang}.},label={code:jqlang}]
static const char* jvp_literal_number_literal(jv n) 
{   ...
    if (plit->literal_data == NULL) {
    int len = jvp_dec_number_ptr(n)->digits + 14;
    plit->literal_data = jv_mem_alloc(len);
    decNumberToString(pdec, plit->literal_data);
    }
    ... 
}
\end{lstlisting}

\textbf{\emph{Segmentation Fault in bash and mksh.}} 
During testing of both bash and mksh, we encounter segmentation fault errors. In bash, the issue occurred in \texttt{parse.y} within the function \texttt{pop\_string}, where accessing \texttt{t->expander->flags \&= \textasciitilde AL\_BEINGEXPANDED} (Line 21 in Code~\ref{code:bash}) failed due to inaccessible addresses set for variables \texttt{t->expand} and \texttt{t->next}. Similarly, in mksh, in \texttt{tree.c}, the function \texttt{wdscan}'s first parameter, \texttt{wp}, pointed to an inaccessible address (Line 7 in Code~\ref{code:mksh}). The reasons behind these issues in both applications are currently unclear. We have provided proof-of-concept (POC) exploits for both bugs to the developers and are awaiting their responses.

\begin{lstlisting}[language=C, emphstyle={\color{blue}}, caption={Segmentation fault in \texttt{bash}.},label={code:bash}]
static void
pop_string ()
{
  STRING_SAVER *t;
  FREE (shell_input_line);
  shell_input_line = pushed_string_list->saved_line;
  shell_input_line_index = pushed_string_list->saved_line_index;
  shell_input_line_size = pushed_string_list->saved_line_size;
  shell_input_line_len = pushed_string_list->saved_line_len;
  shell_input_line_terminator = pushed_string_list->saved_line_terminator;
#if defined (ALIAS)
  if (pushed_string_list->expand_alias)
    parser_state |= PST_ALEXPNEXT;
  else
    parser_state &= ~PST_ALEXPNEXT;
#endif
  t = pushed_string_list;
  pushed_string_list = pushed_string_list->next;
#if defined (ALIAS)
  if (t->expander)
    t->expander->flags &= ~AL_BEINGEXPANDED;
#endif
  free ((char *)t);
  set_line_mbstate ();
}
\end{lstlisting}

\begin{lstlisting}[language=C, emphstyle={\color{blue}}, caption={Segmentation fault in \texttt{mksh}.},label={code:mksh}]
const char *
wdscan(const char *wp, int c)
{
    int nest = 0;
    
    while (/* CONSTCOND */ 1)
        switch (*wp++) {
        case EOS:
            return (wp);
        case ADELIM:
            if (c == ADELIM && nest == 0)
                return (wp + 1);
            if (ord(*wp) == ORD(/*{*/ '}'))
                goto wdscan_csubst;
            /* FALLTHROUGH */
    ...
    }
}
\end{lstlisting}

\end{document}